\documentclass[namedreferences]{SolarPhysics}
\usepackage[optionalrh]{spr-sola-addons} 
\usepackage{graphicx}                    
\usepackage{color}                       
\usepackage{url}                         
\graphicspath{{./figs/}}

\usepackage{amsmath}

\newcommand{\eg}{e.g.}

\begin{document}

\begin{article}

\begin{opening}

\title{Reconstructing the 3-D Trajectories of CMEs in the Inner Heliosphere}

%
\author{Shane A.~\surname{Maloney}$^{1}$\sep
		Peter T.~\surname{Gallagher}$^{1}$\sep
		R. T. James~\surname{McAteer}$^{1}$\sep
	}

%

%
\institute{$^{1}$School of Physics, Trinity College Dublin, Dublin 2, Ireland.
		email: \url{maloneys@tcd.ie} 
	}

\begin{abstract}
A method for the full three-dimensional (3-D) reconstruction of the trajectories of coronal mass ejections (CMEs) using {\it Solar TErrestrial Relations Observatory} (STEREO) data is presented. Four CMEs that were simultaneously observed by the inner and outer coronagraphs (COR1 and 2) of the Ahead and Behind STEREO satellites were analysed. These observations were used to derive CME trajectories in 3-D out to $\sim$15\,R$_{\odot}$. The reconstructions using COR1/2 data support a radial propagation model. Assuming pseudo-radial propagation at large distances from the Sun (15\,--\,240\,R$_{\odot}$), the CME positions were extrapolated into the Heliospheric Imager (HI) field-of-view. We estimated the CME velocities in the different fields of view. It was found that CMEs slower than the solar wind were accelerated, while CMEs faster than the solar wind were decelerated, with both tending to the solar wind velocity.
\end{abstract}

%

\end{opening}

%

%




\section{Introduction}
	\label{S-intro}
 	
Coronal mass ejections (CMEs) are large scale eruptions of plasma and magnetic field which propagate from the Sun into the Heliosphere. A typical CME has a magnetic field strength of $\sim$50\,nT, a mass in the range of $10^{13}$--$10^{16}$\,g \cite{Vourlidas:2002p3006} and velocity between $\sim$10 and $\geq$2500\,km\,s$^{-1}$ close to the Sun \cite{Gopalswamy:2004p3007}. At 1\,AU, CMEs velocities (300--1000\,km\,s$^{-1}$) tend to be closer to the solar wind speed \cite{Lindsay:1999p3850,Wang:2005p19,Gopalswamy:2007p18}. Although CMEs often exhibit a three-part structure which consists of a bright front followed by a dark cavity and bright core, they may also exhibit more complex structures (\eg, \citeauthor{Pick:2006p30}\citeyear{Pick:2006p30}).

The full dynamical evolution of CMEs includes three phases; initiation, acceleration and propagation \cite{Zhang:2001p452}. The initiation  phase involves a slow build up of energy through various mechanisms, such as shearing (\eg, \opencite{Mikic:1994p898M}; \citeauthor{Amari:1996p913A}\citeyear{Amari:1996p913A}) or flux emergence and flux cancellation (\eg, \citeauthor{Amari:200049A}\citeyear{Amari:200049A}; \citeauthor{Krall:2001p26}\citeyear{Krall:2001p26}). A number of models, such as the ``magnetic breakout'' model \cite{Antiochos:1999p17,Lynch:2008p1403} and the flux rope model \cite{Chen:1996p27499}, have been developed which predict both dynamics and observable features during both the initiation and acceleration phases (see \citeauthor{Forbes:2006p43}\citeyear{Forbes:2006p43} for a recent review). During the propagation phase, the bulk of the acceleration has ceased, but {\it in situ} measurements at 1\,AU show CMEs tend towards the solar wind velocity. This residual acceleration has been modelled in the context of a ``drag'' force, for example, the ``snow plough'' model \cite{Tappin:2006p47} and aerodynamic drag model \cite{Cargill:2004p12}.

During the {\it SOlar and Heliospheric Observatory} (SOHO) era CMEs could only be routinely observed in white light up to $\sim$32\,R$_{\odot}$ using the Large Angle Spectrometric Coronagraph (LASCO; \citeauthor{Brueckner:1995p3300}\citeyear{Brueckner:1995p3300}), and sometimes to Earth using the Solar Mass Ejection Imager (SMEI; \citeauthor{Jackson:2004p2417}\citeyear{Jackson:2004p2417}; \citeauthor{Howard:2007p51}\citeyear{Howard:2007p51}). Shocks driven by CMEs are observed in radio data, which rely on model densities to relate the observations to heights in the corona and provide no information on the direction of propagation. Some work has been done using radio data from two observatories to reconstruct three dimensional (3-D) positions from radio triangulation (for example \citeauthor{Reiner:1998p1923}\citeyear{Reiner:1998p1923} and \citeauthor{Renier:2007pA2}\citeyear{Renier:2007pA2}). Another possibility to track CMEs over large distances is during fortunate quadrature observations such as SOHO-Sun-{\it Ulysses} \cite{Bemporad:2003p1934}. CMEs have also been tracked out into the Heliosphere using {\it in situ} measurements from various spacecraft such as {\it Wind}, ACE, {\it Ulysses}, {\it Cassini}, {\it Voyager} 1 and 2 (\citeauthor{Lario:2005p3847}\citeyear{Lario:2005p3847}, \citeauthor{Richardson:2005p3848}\citeyear{Richardson:2005p3848}, \citeauthor{Gopalswamy:2005p3849}\citeyear{Gopalswamy:2005p3849} and references therein). \citeauthor{Howard:2008p3598} (\citeyear{Howard:2008p3598}) used a triangulation method on observations from SOHO/LASCO and the {\it Solar TErrestrial RElations Observatory} (STEREO) spacecrafts to obtain CME positions in 3-D. To identify events over interplanetary distance scales, when neither radio or quadrature observations are available, it is necessary to compare the the coronagraph images with {\it in situ} data. The relation between images and the {\it in situ} data is complex: Coronagraphs image solar radiation which has been Thomson scattered by electrons in the CME; {\it in situ} measurements provide the actual densities and magnetic field of  the part of the CME which passes over the instrument (flank or nose). All of  this is compounded when multiple events occur in a short time frame. Also, coronagraphs produce images that are subject to projection effects which is one of the leading sources of uncertainty in determining kinematics \cite{Howard:2008p15}.

Observations from one perspective only have information about two of the three spatial dimensions. This has serious implications for many solar observations. For example, in extreme ultraviolet (EUV) single perspective images, the tilt and height of loops is ambiguous and therefore the loop length and scale heights can not be determined accurately \cite{Aschwanden:2008p249}. In coronagraph images, only the component of height in the plane of sky is measured. 
Using stereoscopy techniques the reconstruction of the 3-D coordinates of features identified in images from two vantage points is a well defined linear problem \cite{Inhester:2006p2249}. The biggest challenge in using this method is the identification and matching of features in stereo image pairs. Tracking the true 3-D trajectories of CMEs will facilitate comparison between observations and theory, and is necessary in order to better understand CME acceleration and propagation. From a space weather perspective, full 3-D reconstruction of CME trajectories will enable more accurate estimates of CME arrival times at 1 AU (\citeauthor{Owens:2004p1402} \citeyear{Owens:2004p1402}).

The launch of the twin STEREO spacecraft in December 2006 has provided such dual vantage point observations of the Sun allowing 3-D reconstruction of solar features. 
The two perspectives of the spacecraft allow for the stereoscopic reconstruction of features in EUVI \cite{Aschwanden:2008p249}, and the same 3-D reconstruction can be applied to the COR1/2 images to provide 3-D data out to 15\,$R_{\odot}$. By assuming that CMEs follow the same trajectories in the $x-y$ plane (ecliptic plane) in the HI field-of-view (FOV) as in the COR1/2 FOV, the true position of the CME may be tracked from the Sun out to $\sim$1\,AU and beyond. This addresses the effects of projection in coronagraph images (which is even more important in the HI FOV), and the lack of consistent observations in the inner Heliosphere \citeauthor{Owens:2004p1402} \citeyear{Owens:2004p1402}).

In Section \ref{S-Obsred}, the observations and their reduction and analysis are discussed. The 3-D reconstruction methods for the COR and HI instruments are described in Section \ref{S-recon}. The extrapolation of the coordinates into the HI FOV using the COR data is also explained in this section, while the Appendix contains the details of the coordinate transformations for the HI and COR instruments. In Section \ref{S-results}, of four sample events are presented. Section \ref{S-disc} contains a discussion of the results, the conclusions drawn, and future work.

\section{Observations and Data Reduction }
	\label{S-Obsred}
	\subsection{Observations}
 	\label{SS-Obs}
In this paper we present a sample of four CMEs observed by the STEREO/SECCHI instrument suites on both the Ahead (A) and Behind (B) spacecraft. The dates of the observations are 8-13 October 2007, 25-28 March 2008, 9-10 April 2008 and 9-13 April 2008. 
	
The inner SECCHI coronagraph, COR1, is a Lyot internally occulted refractive telescope which images the white light corona from 1.4--4\,$R_{\odot}$ with 7.5 arcsec pixel$^{-1}$ plate scale \cite{Howard:2008p67}. The COR2 instrument is an externally occulted Lyot coronagraph, of similar design to that of the SOHO/LASCO C2 and C3 coronagraphs, which images the corona from 2.5--15\,R$_{\odot}$ with 14.7 arcsec pixel$^{-1}$ \cite{Howard:2008p67}. Both COR1 and COR2 take sequences of three polarised images at $0^{\circ}$ and $\pm60^{\circ}$ degrees for a complete polarisation sequence. This results in a cadence of 8 and 15 minutes respectively. These images are telemetered to the ground where they can be combined to give both total brightness (B) and polarised brightness (pB) images.

The HI instrument is a combination of two refractive optical telescopes (HI1 \& HI2) with a multi-vane, multi stage light rejection system. HI1 has a FOV of 20$^{\circ}$ degrees centred on an elongation of 13.28$^{\circ}$\,degrees with $70$\,arcsec pixel$^{-1}$ plate scale, while HI2 has a FOV of 70$^{\circ}$\,degrees centred on an elongation of 53.36$^{\circ}$ degrees from the Sun with $4$ arcmin pixel$^{-1}$ plate scale. The HI baffle system provides extremely high stray light rejection levels of $3\times10^{-13}\,$B$_{\odot}$ and $3\times10^{-14}\,$B$_{\odot}$ for HI1 and HI2 which is needed to image the CMEs. In order to attain the required signal-to-noise ratio needed to image the extremely faint CMEs, the HIs take long exposure images. Due to limited dynamic range, and to avoid cosmic ray swamping, a series of shorter exposures are taken and summed on-board: In the case of HI1, this means that 30 images of 40 second exposure are summed (total exposure time 1200\,s) to give images at a cadence of 40 minutes; for HI2 99 images with a 50 second exposure time are summed (total exposure time 4950\,s) leading to an image cadence of 2 hours. As neither HI1 or HI2 have shutters, all images must be corrected for the smearing caused by shutterless readout (each row sees the scene from the pixel below it). Finally some of the objects in the HI FOV are bright enough to saturate the CCD, these bleed along columns which allows them to be identified later \cite{Eyles:2009p193}.

 	\subsection{Data Reduction}
 	\label{SS-red}
The data were reduced and calibrated using ${\it secchi\_prep.pro}$ from the SECCHI tree within the {\it SolarSoft} library \cite{Freeland:1998p3546}. This consisted of de-biasing and application of flat-field corrections for all data. For both COR1 and COR2 data, additional corrections for exposure time, vignetting and an optical distortion were carried out. Furthermore, for COR1, a model background was subtracted to remove the static corona. The HI data required additional corrections for the smearing and pixel bleeding (Section \ref{SS-Obs}). Columns with bleeding were and then replaced with a known value. Finally, the pointing information in the HI FITS headers were updated with pointing information derived from the star fixes in the images \cite{Brown:2008p181}. All of the images were left in units of digital numbers, with the exception of the HI data which is implicitly converted to digital numbers per second when the shutterless correction is applied.

 	\subsection{Data Processing}
 	\label{SS-proces}
The data were processed further to enhance the faint transient CME features. The COR1 and COR2 images were processed in a similar manner. First the data were clipped so that all non-coronal pixels (occulter and edge of the CCD) are set to a known value. The images were also normalised by dividing each image by its mean to reduce image-to-image intensity variation and facilitate production of better running difference images. After this, a further background was created by taking the median of the data cube along the time axis and subtracted to remove more semi-stationary coronal features. A running difference sequence was then made by subtracting the previous image from the current one. This was then median filtered and re-scaled.
	
For HI, different processing methods were applied as a typical CME is only $\sim$\,1$\%$ of the natural background of the images. The star fields in the FOV of the instrument can be approximated as constant over an exposure time, but there will be a shift due to the motion of the satellite. If the offset between two subsequent images can be found, then the images can be shifted and subtracted and only transient features will be left
\begin{equation}
	\label{Eq-aa}
	\Delta I \left (x, y \right ) _{i} = I \left (x, y \right )_{i} - I \left (x+\alpha, y+\beta \right )_{i-1} \\
 \end{equation} where $\alpha$ and $\beta$ are the  $x$ and $y$ offsets found by the cross correlation. However, the presence of a geometrical optical distortion in the HI images makes this difficult. A method to approximate this offset was developed . This involved extracting a region (512\,$\times$\,512) at the image centre, where the distortion is at a minimum, and performing a local cross-correlation in a running window (128\,$\times$\,128) around this sub-image. The average offset from this procedure was used to shift the image. The resulting running difference images were then median filtered. 
	\begin{figure}    
		\centerline{\hspace*{-0.015\textwidth}
              		\includegraphics[width=0.9\textwidth,clip=]{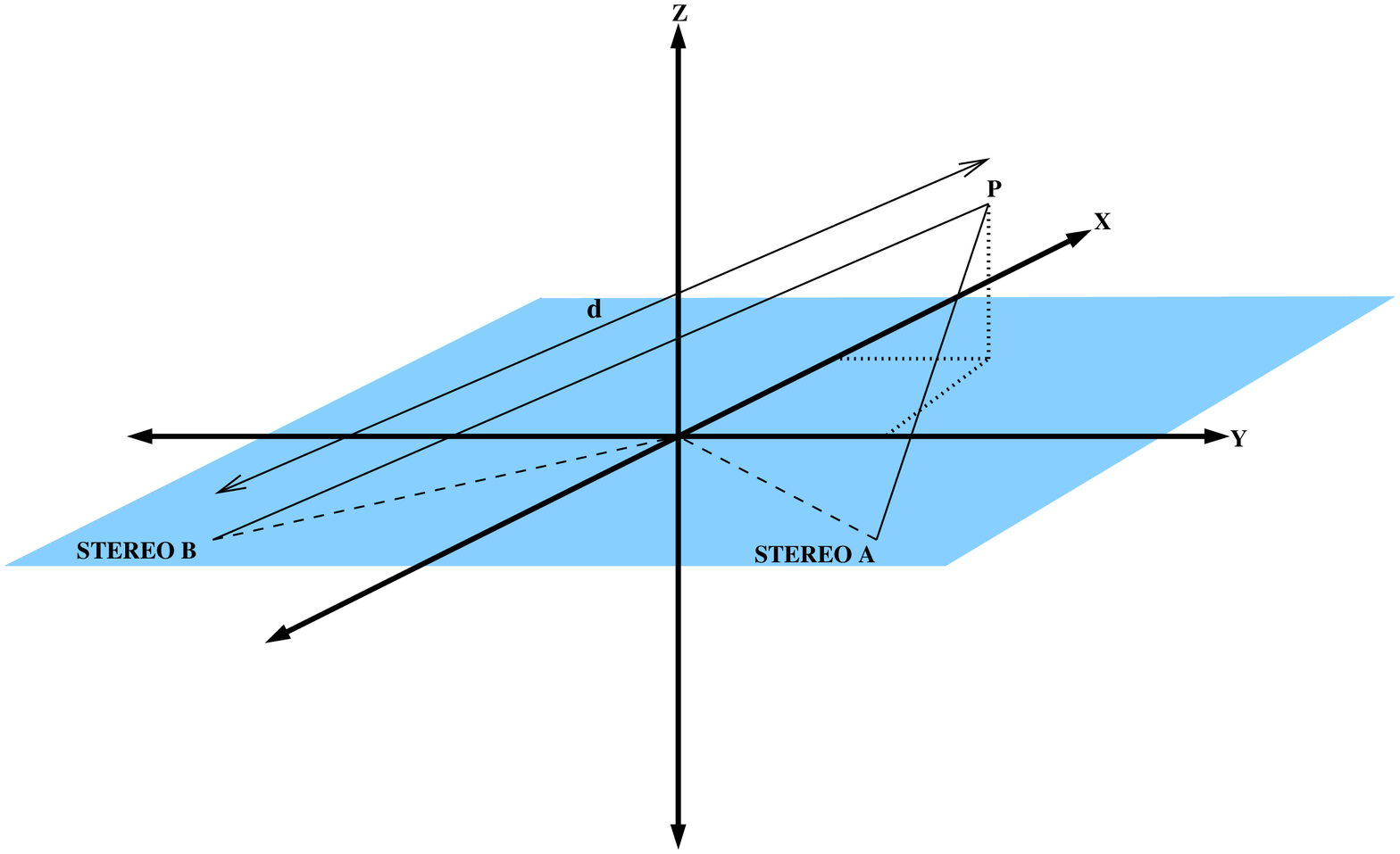}
		}
          \centerline{\hspace*{-0.015\textwidth}
          			\includegraphics[width=0.8\textwidth,clip=]{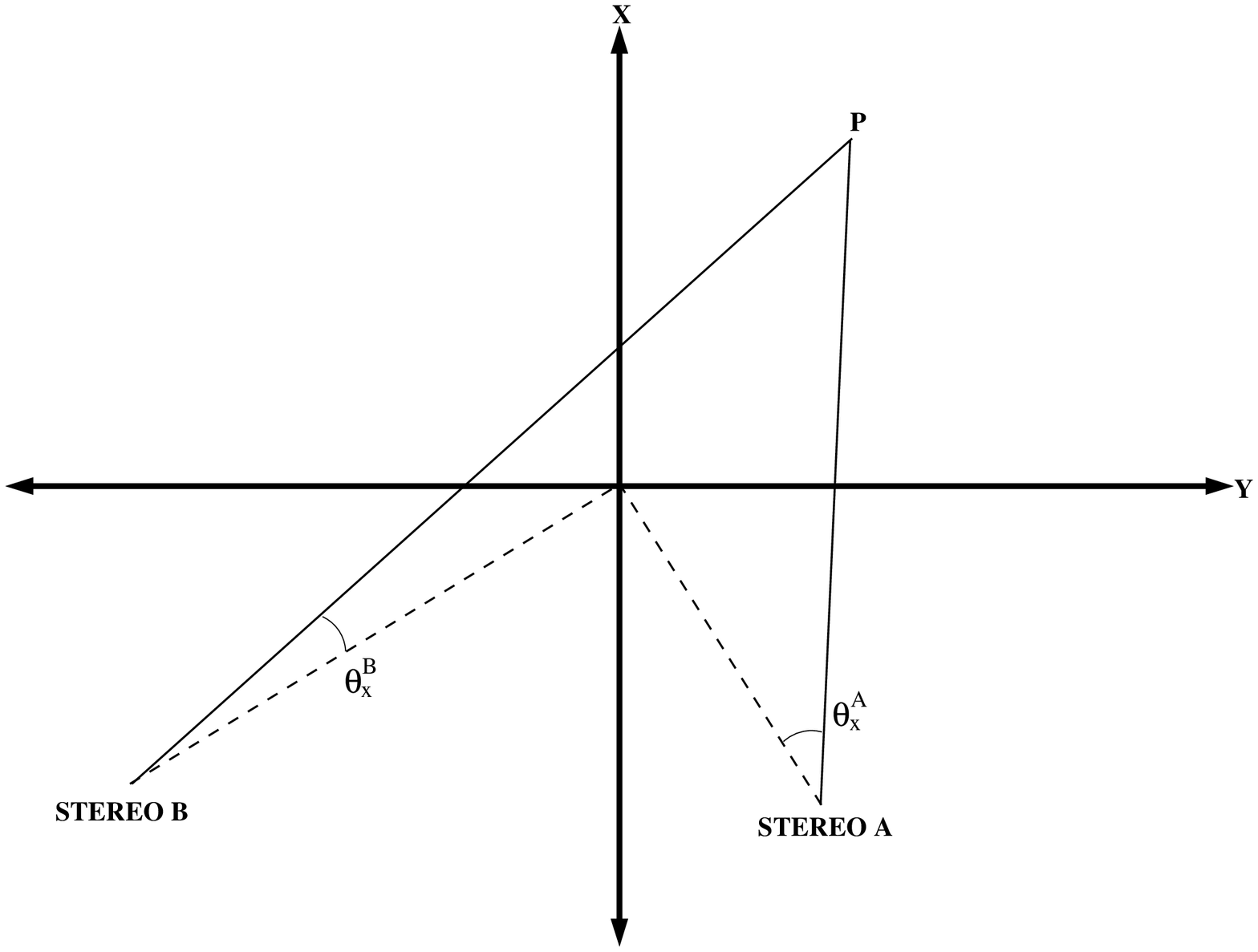}
         	}
		\caption{{\it Top:} 3-D representation of the geometry of the observations with the lines-of-sight (LOS) from the two spacecraft (solid), spacecraft Sun centre lines (dashed) and the point to be reconstructed P. {\it Bottom:} Projection in the $x-y$ plane showing the angles $\theta_{x}^{A}$ and $\theta_{x}^{B}$, where $x$ corresponds to image pixel axis. The point P is found by solving the equations of the LOS for the their intercept. The same can be done in the $x-z$ plane for $\theta_{y}$.}
    		\label{F-3drecon}
\end{figure}

	\begin{figure}    
		\centerline{\hspace*{0.015\textwidth}
              		\includegraphics[width=0.5\textwidth,clip=]{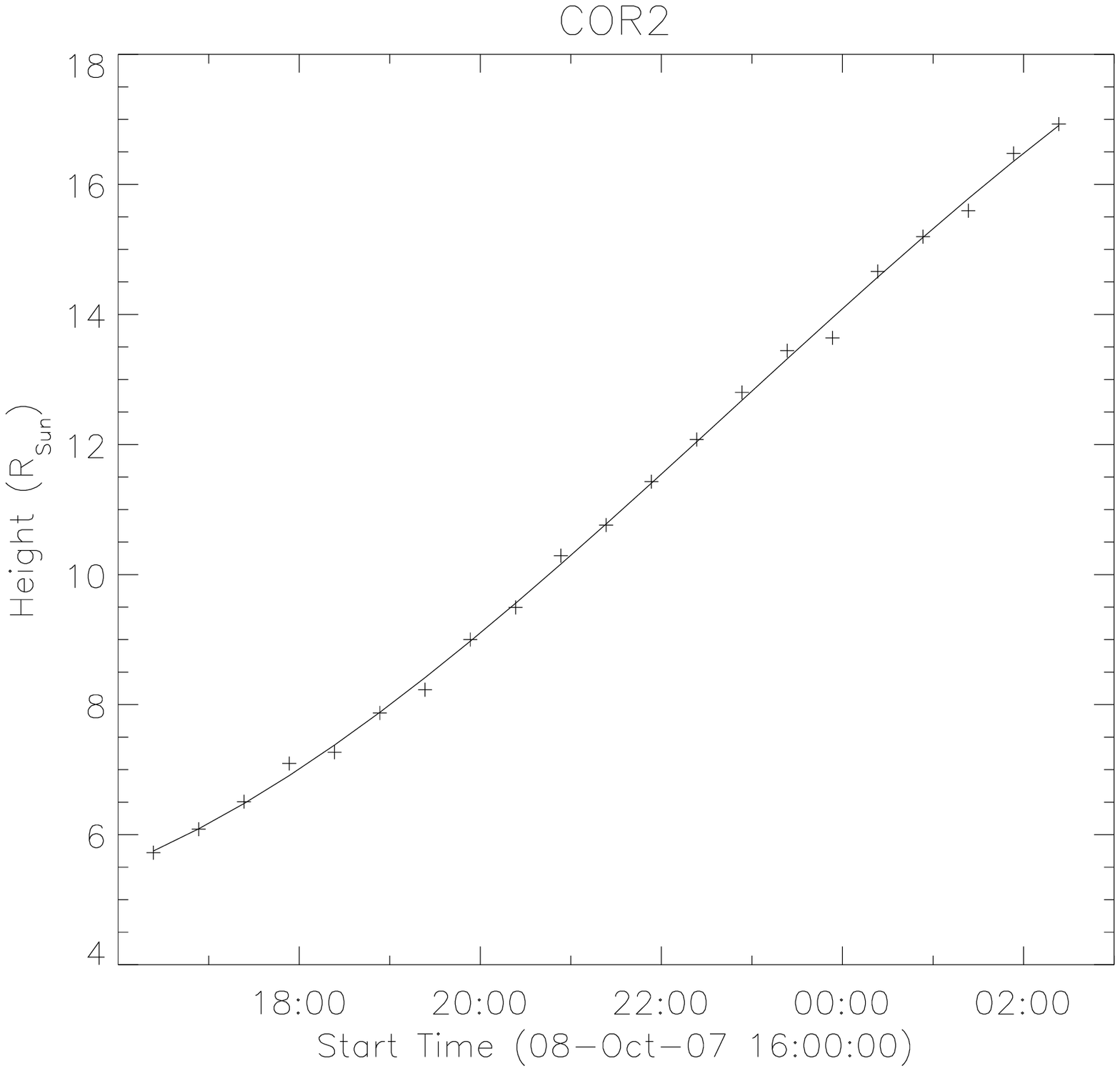}
              		\hspace*{-0.03\textwidth}
              		\includegraphics[width=0.5\textwidth,clip=]{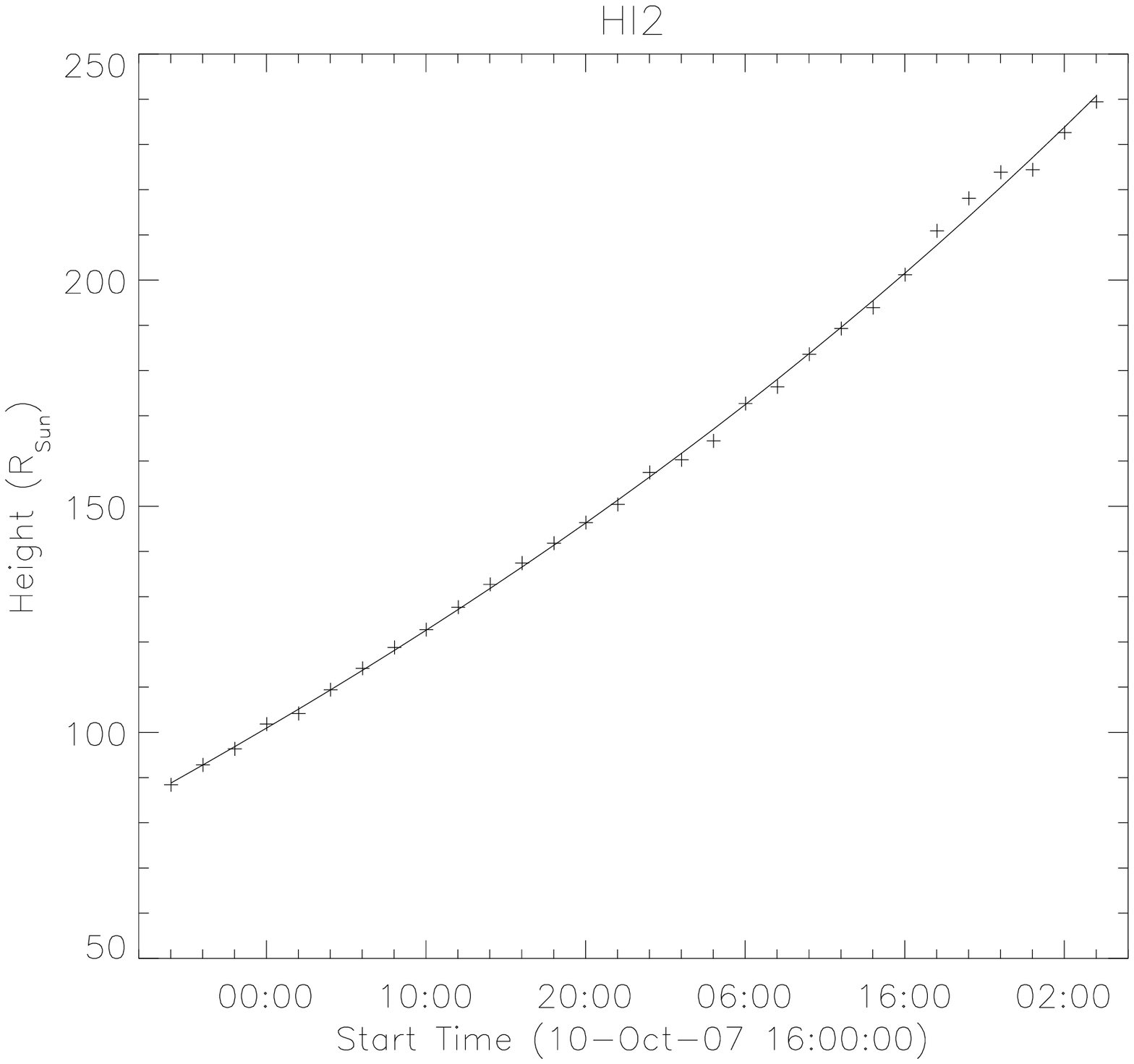}	              		
		}
     		\vspace{-0.35\textwidth}   
     		\centerline{\Large \bf     
        	\hfill}
     		\vspace{0.31\textwidth}    
		 \centerline{\hspace*{0.015\textwidth}
		 		\includegraphics[width=0.5\textwidth,clip=]{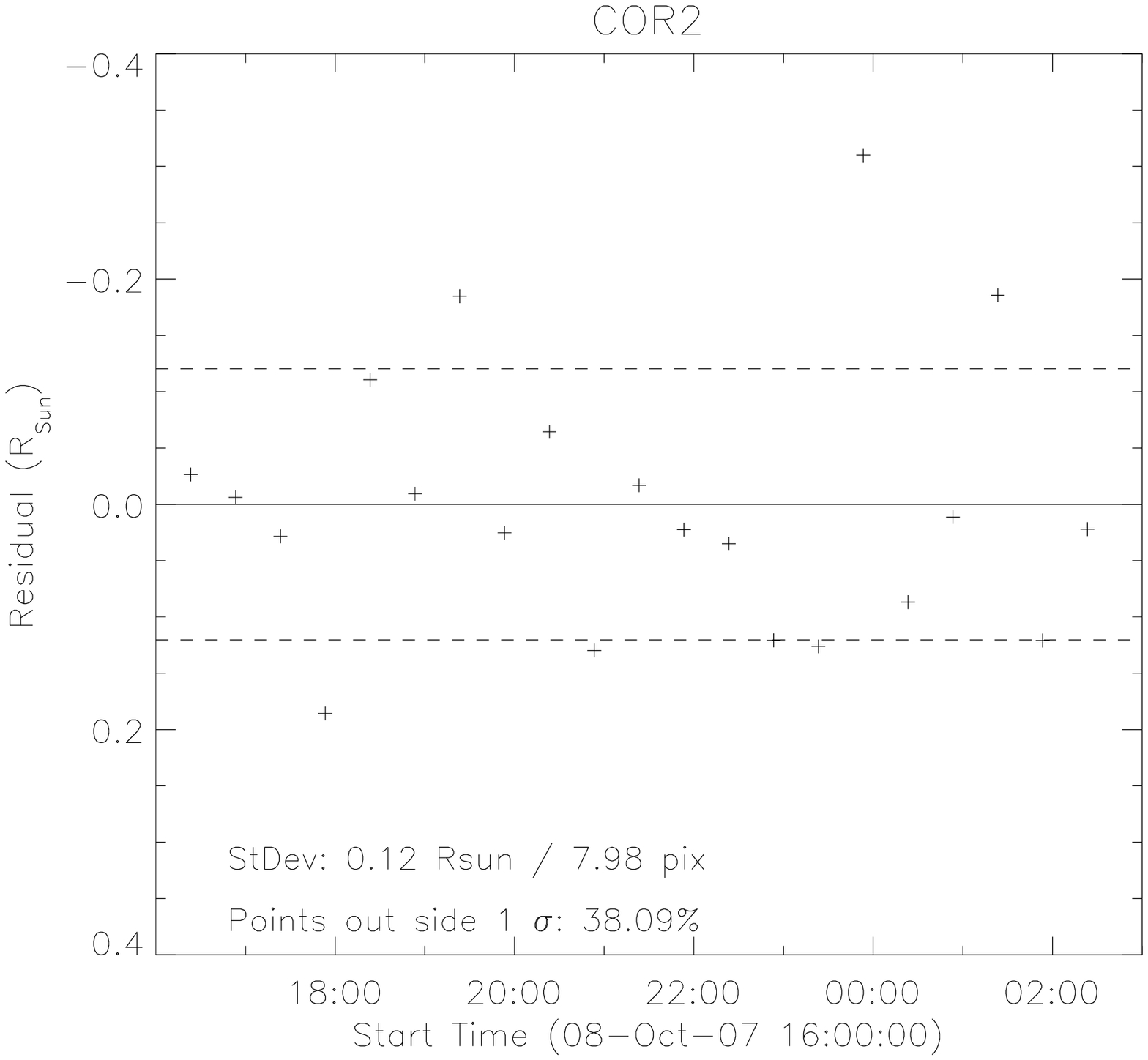}
              			\hspace*{-0.03\textwidth}
              			\includegraphics[width=0.5\textwidth,clip=]{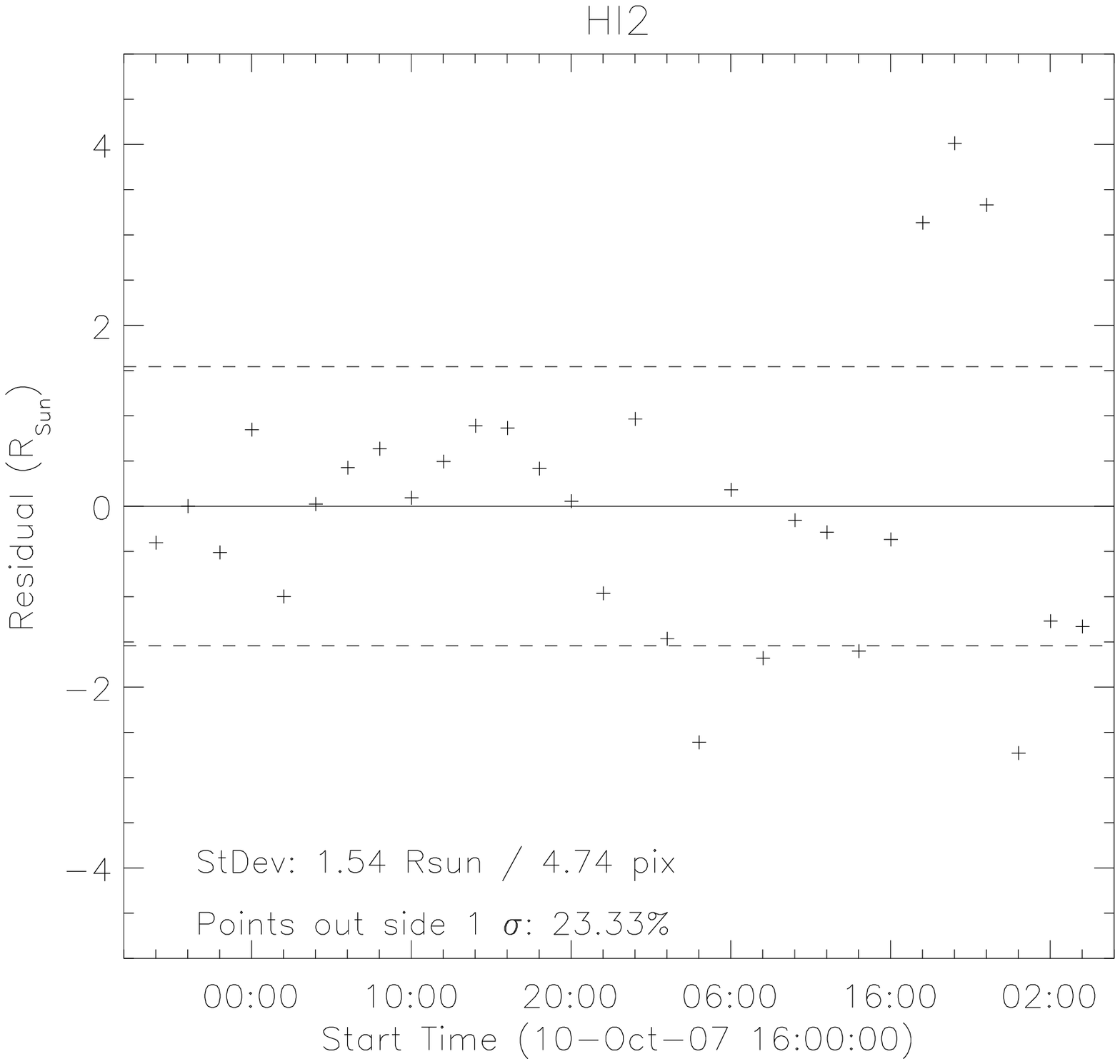}    
		              }
		     \vspace{-0.35\textwidth}   
		     \centerline{\Large \bf     
		         \hfill}
		     \vspace{0.31\textwidth}    
		\caption{{\it Top row:} Height versus time data for the 2007 8-13 October event in the COR2 and HI2 FOV, the line is the best fit to the data with a 3\textsuperscript{rd} order polynomial. {\it Bottom row:} The de-trended data. The standard deviation $\sigma$ and the percentage of points out side of a one $\sigma$ error are shown. The dashed lines are plotted at $\pm$\,1\,$\sigma$.}
    		\label{err3}
	\end{figure}

%
 \begin{figure}    
 	\centerline{\includegraphics[width=0.95\textwidth,clip=]{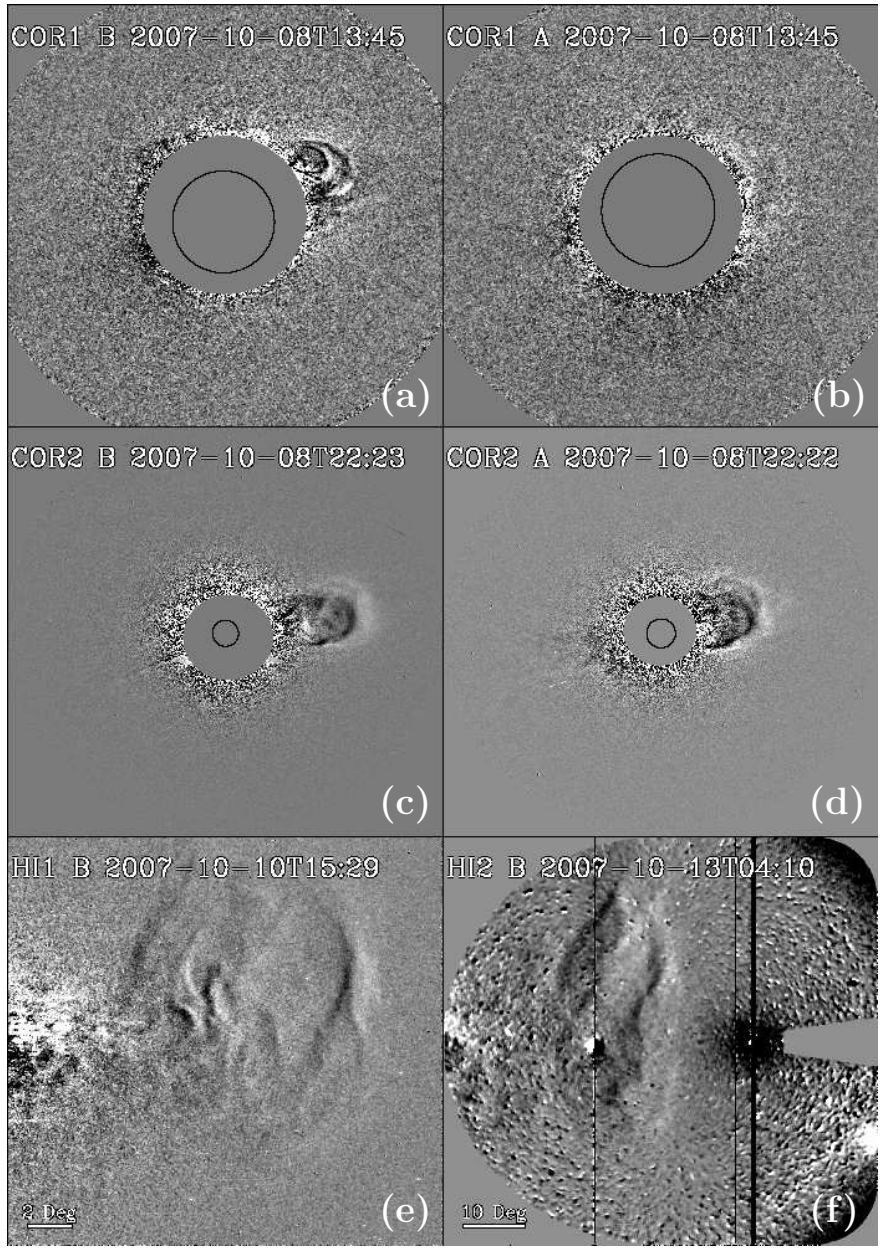}}
	\vspace{-0.95\textwidth}   
		     \centerline{\Large \bf     
		     \hspace{0.415 \textwidth} \color{white}{(a)}
		     \hspace{0.38\textwidth}  \color{white}{(b)}
	\hfill}
	 \vspace{0.40\textwidth}
	 	     \centerline{\Large \bf     
		     \hspace{0.415 \textwidth} \color{white}{(c)}
		     \hspace{0.38\textwidth}  \color{white}{(d)}
		     \hfill}
	\vspace{0.40\textwidth}
		 	\centerline{\Large \bf     
		     \hspace{0.415 \textwidth} \color{white}{(e)}
		     \hspace{0.38\textwidth}  \color{white}{(f)}
		     \hfill}
	\vspace{0.02\textwidth}    
	 \caption{Sample images from from the 8-13 October 2007 event. {\bf (a)} COR1 Behind 13:45 UT.  {\bf (b)} COR1 Ahead 13:45 UT. {\bf (c)} COR2 Behind 22:23 UT. {\bf (d)} COR2 Ahead 22:22 UT. {\bf (e)} HI1 Behind 2007 October 10 15:29 UT. {\bf (f)} HI2 Behind 2007 13 October 04:10 UT. The Sun is to the left in both of the HI images.}
   	\label{F-20071008mimg}
\end{figure}

%
 \begin{figure}    
 	\centerline{\includegraphics[width=0.95\textwidth,clip=]{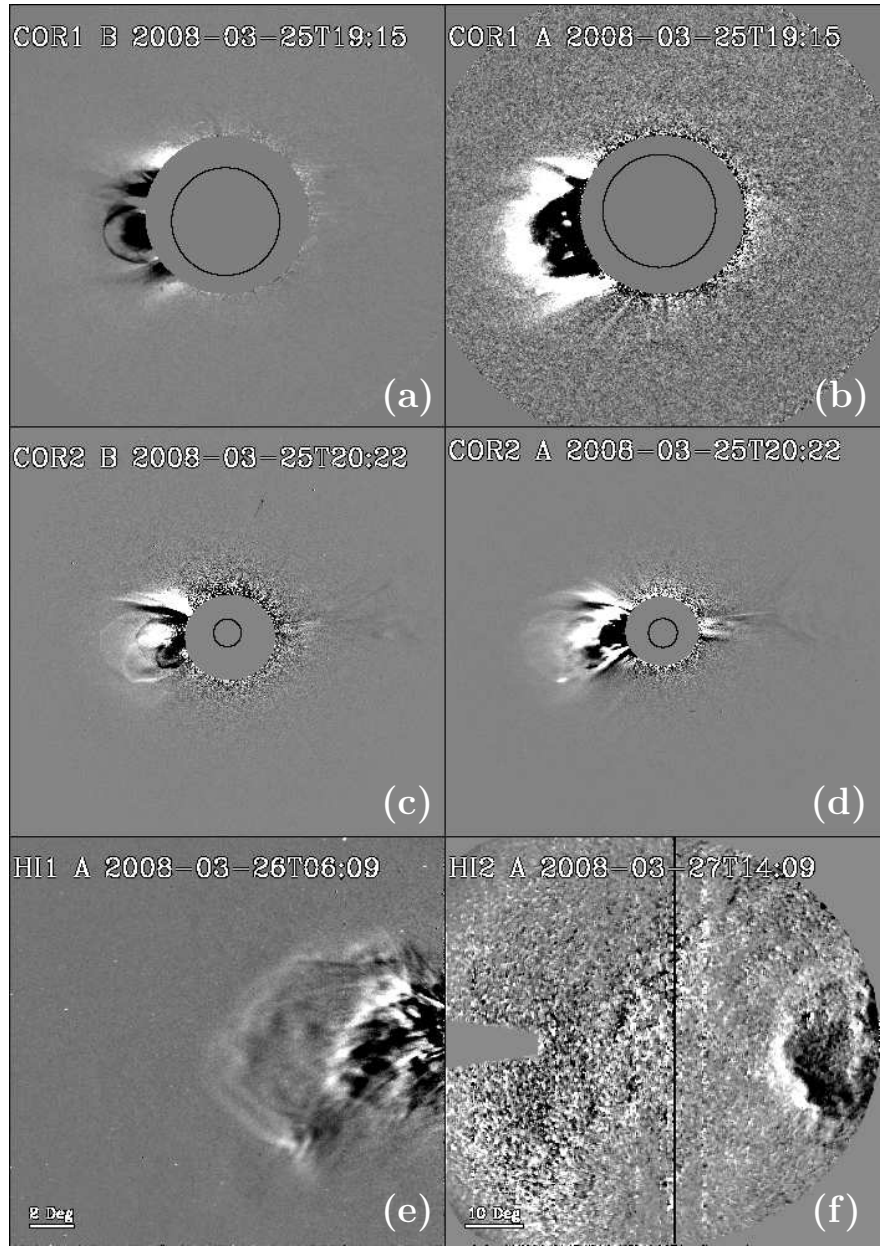}}
		\vspace{-0.95\textwidth}   
		     \centerline{\Large \bf     
		     \hspace{0.415 \textwidth} \color{white}{(a)}
		     \hspace{0.38\textwidth}  \color{white}{(b)}
	\hfill}
	 \vspace{0.40\textwidth}
	 	     \centerline{\Large \bf     
		     \hspace{0.415 \textwidth} \color{white}{(c)}
		     \hspace{0.38\textwidth}  \color{white}{(d)}
		     \hfill}
	\vspace{0.40\textwidth}
		 	\centerline{\Large \bf     
		     \hspace{0.415 \textwidth} \color{white}{(e)}
		     \hspace{0.38\textwidth}  \color{white}{(f)}
		     \hfill}
	\vspace{0.02\textwidth}  
	 \caption{Sample images from from the 25-28 March 2008 event. {\bf (a)} COR1 Behind 19:15. {\bf (b)} COR1 Ahead 19:15 UT. {\bf (c)} COR2 Behind 20:22 UT. {\bf (d)} COR2 Ahead 20:22 UT. {\bf (e)} HI1 Behind 2008 March 26 06:09 UT. {\bf (f)} HI2 Behind 2008 March 27 14:09 UT. The Sun is to the right in both of the HI images.}
   	\label{F-20080325mimg}
\end{figure}

%
 \begin{figure}    
 	\centerline{\includegraphics[width=0.95\textwidth,clip=]{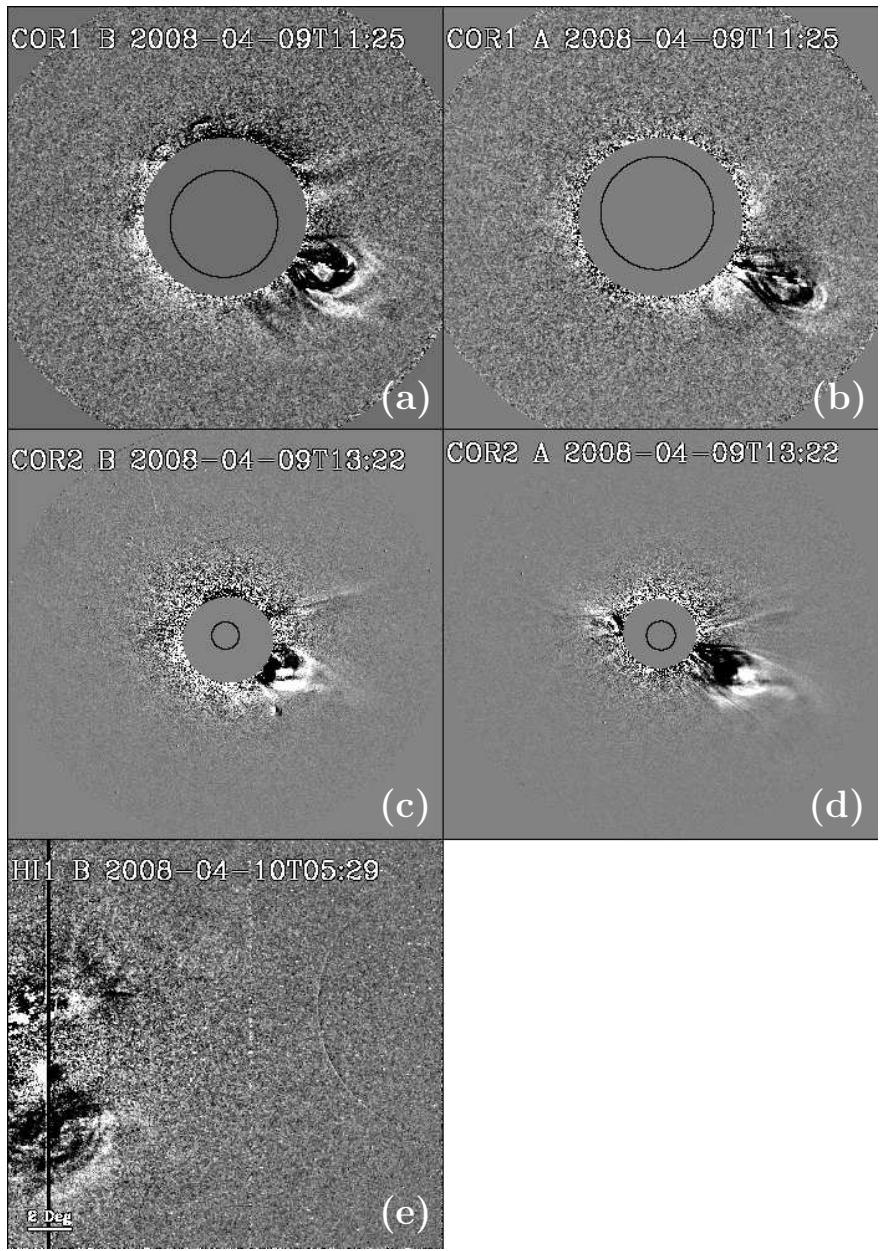}}
		\vspace{-0.95\textwidth}   
		     \centerline{\Large \bf     
		     \hspace{0.415 \textwidth} \color{white}{(a)}
		     \hspace{0.38\textwidth}  \color{white}{(b)}
	\hfill}
	 \vspace{0.40\textwidth}
	 	     \centerline{\Large \bf     
		     \hspace{0.415 \textwidth} \color{white}{(c)}
		     \hspace{0.38\textwidth}  \color{white}{(d)}
		     \hfill}
	\vspace{0.40\textwidth}
		 	\centerline{\Large \bf     
		     \hspace{0.415 \textwidth} \color{white}{(e)}
		     \hspace{0.38\textwidth}  \color{white}{(f)}
		     \hfill}
	\vspace{0.02\textwidth}  
	 \caption{Sample images from the 9-10 April 2008 event. {\bf (a)} COR1 Behind 11:25 UT. {\bf (b)} COR1 Ahead 11:25 UT. {\bf (c)} COR2 Behind 13:22. {\bf (d)} COR2 Ahead 13:22 UT. {\bf (e)} HI1 Behind 2008 April 10 05:29 UT. The Sun is to the left in the HI image.}
   	\label{F-20080409mimg}
\end{figure}

%
 \begin{figure}    
 	\centerline{\includegraphics[width=0.95\textwidth,clip=]{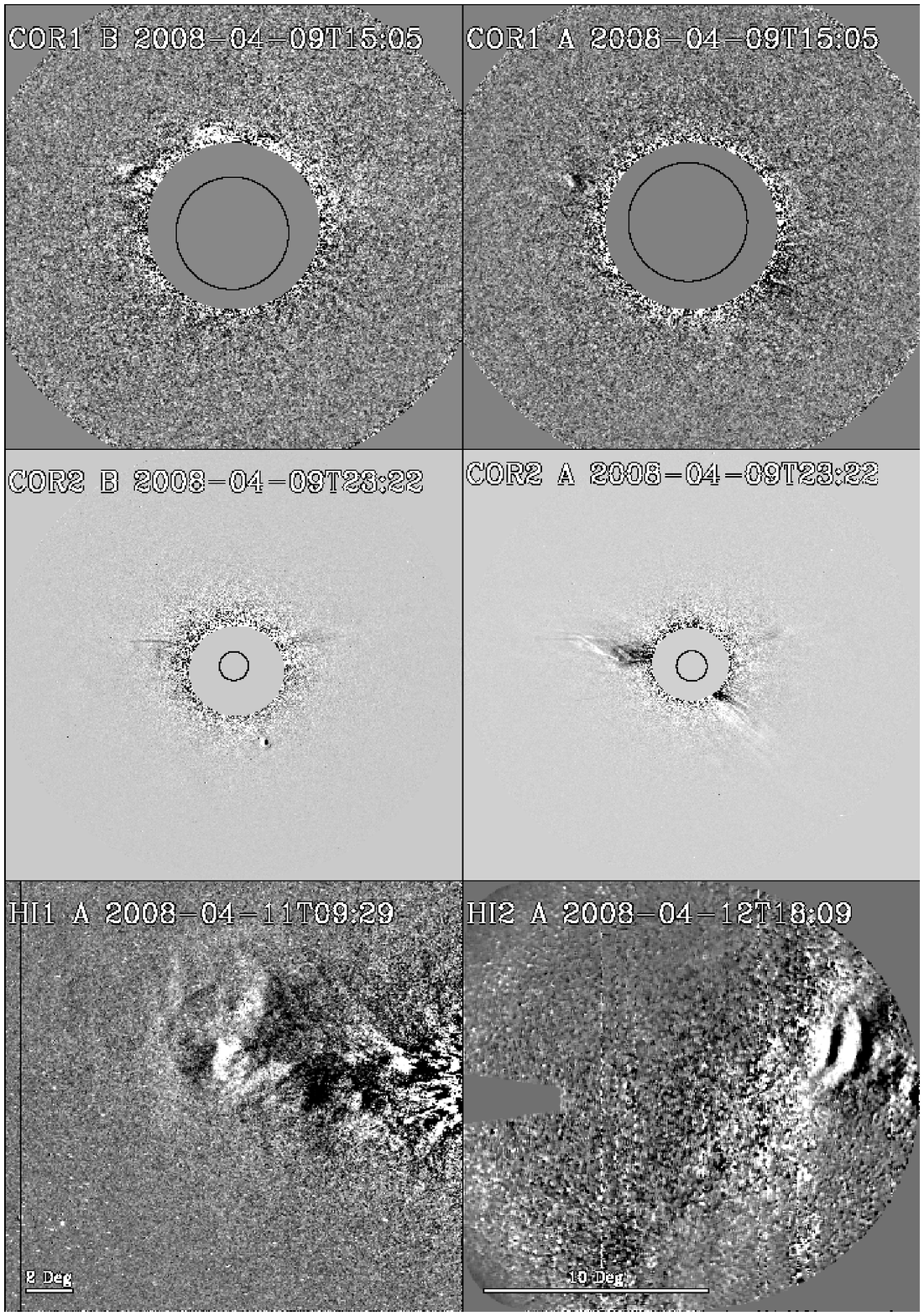}}
		\vspace{-0.95\textwidth}   
		     \centerline{\Large \bf     
		     \hspace{0.415 \textwidth} \color{white}{(a)}
		     \hspace{0.38\textwidth}  \color{white}{(b)}
	\hfill}
	 \vspace{0.40\textwidth}
	 	     \centerline{\Large \bf     
		     \hspace{0.415 \textwidth} \color{white}{(c)}
		     \hspace{0.38\textwidth}  \color{white}{(d)}
		     \hfill}
	\vspace{0.40\textwidth}
		 	\centerline{\Large \bf     
		     \hspace{0.415 \textwidth} \color{white}{(e)}
		     \hspace{0.38\textwidth}  \color{white}{(f)}
		     \hfill}
	\vspace{0.02\textwidth}  
	 \caption{Sample images from the 9-13 April 2008 event. {\bf (a)} COR1 Behind 15:05 UT. {\bf (b)} COR1 Ahead 15:05 UT. {\bf (c)} COR2 Behind 23:22 UT. {\bf (d)} COR2 Ahead 23:22 UT. {\bf (e)} HI1 Behind 2008 April 11 09:29 UT, {\bf (f)} HI2 Behind 2008 April 12 18:09 UT. The Sun is to the right in both of the HI images.}
   	\label{F-20080410mimg}
\end{figure}

\section{3-D Reconstruction}
	\label{S-recon}

The reconstruction of 3-D information from images of an object observed from two different perspectives is well studied and the general reconstruction methods are well developed. The technique used in this study relies on the epipolar constraint which states that any feature identified on an epipolar line in one stereo image must lie along the same epipolar line in the other stereo image \cite{Inhester:2006p2249}. An epipolar line is the projection of the epipolar plane (the plane containing the two observers STEREO A, B and the point to be reconstructed, P; see top panel of Figure \ref{F-3drecon}) in the stereo images.

One possible implementation of this method of stereoscopy is outlined below. We implicitly assume that two simultaneous observations of a feature (the point P; see top panel of Figure \ref{F-3drecon}) from two different perspectives are available.  If the feature is identified in one of the images, from A for example, the angles $\theta_{x}^{A}$ and $\theta_{y}^{A}$ can be derived and the only unknown is the the depth of the feature or the distance along the line-of-sight $d$ (see bottom panel of Figure \ref{F-3drecon}). One can arbitrarily pick two values for $d$ thereby giving two points in 3-D space which can then be transformed into Heliocentric-Cartesian coordinates using:
   \begin{eqnarray} 
			& x &= d \, \cos \left ( \theta_{y} \right ) \, \sin \left ( \theta_{x} \right )
			  \label{Eq-xhcc}
			 \\      
			& y &= d \,\sin \left ( \theta_{y} \right )
			 \label{Eq-yhcc}
			 \\  
			& z &= D_{\odot} - d \,\cos \left ( \theta_{y} \right ) \, \cos \left ( \theta_{x} \right )
                       \label{Eq-zhcc}    
   \end{eqnarray} where $d$ is the distance along the line-of-sight, $\theta_{x}^{A}$ and $\theta_{y}^{B}$ are the inclination angles of the line-of-sight with respect to the Sun-spacecraft line in the $x-y$ and $x-z$ planes and $D_{\odot}$ is the distance to the Sun. Finally, these are transformed into a suitable coordinate system for the reconstruction, such as Heliospheric-Earth-Ecliptic (HEE).

The coordinates of this line, as viewed from the second perspective, can then be derived by reversing the procedure and the epipolar line over-plotted on the second image. If the position of the same feature along this epipolar line can be found, this gives all the information needed to complete the reconstruction. The same procedure is carried out for this second point, selecting two values of $d$ and transforming these into the reconstruction coordinate system. Now the problem is reduced to finding the point of intersection of two lines in 3-D, which can be simplified to solving for the intersection point of two lines in 2-D in two planes for example the $x-y$ and $x-z$ plane (see bottom panel of Figure \ref{F-3drecon}).

Unlike the situation in COR1/2, many events in HI FOV have only been observed from one perspective at any given time. The full 3-D coordinates can not be derived with just these observations so there is a need for additional constraints. In the HI FOV, the CMEs are at a large distances from the Sun so we assume that the CME will propagate pseudo-radially, continuing along the trajectory that it followed in the COR1/2 FOV. Based on this assumption, we use the best fit of the trajectory COR1/2 data in the $x-y$ (ecliptic) plane and constrain the CME to propagate along this fit. The equations are solved for the point of intersection between this fit line and the observed line-of-sight yielding the $x$, $y$ position. The $z$ coordinate is then calculated by assuming two distances along the line-of-sight to find the equation of the line with respect to $x$ or $y$, and substituting in relevant coordinate calculated yields the corresponding $z$ value.

The apex of the CME was identified using point-and-click methods. These points were then used to calculate the 3-D coordinates as outlined above (see the Appendix for further details). From these, we calculated the CME launch angles with respect to the ecliptic plane, and in the ecliptic plane with respect to the Sun-Earth line. The heights of the CMEs above the solar surface were then be calculated using $h = (x^2+y^2+z^2)^{1/2}$.
	
The uncertainties were estimated using a statistical approach to find the spread of the data. The 3-D height data is de-trended with a third order polynomial and the moments of the distribution are then calculated. The standard deviation ($\sigma$) of the distribution is taken as a good measure of the uncertainty. As a check on the statistics, the number of data points greater than one $\sigma$ from the mean were calculated and compared to the 31.8\% which we expect if the data is normally distributed. If the number of points in the individual fields-of-view are too small this approach will not work. Figure 2, a sample of images from the 2007 8-13 October event for COR2 and HI2 is shown. The COR2 data is well described by a normal distribution with 38.09\,\% of the data lying outside of one $\sigma$, the outliers are well distributed through the event showing the polynomial fit the data trend well (see left of Figure \ref{err3}). There are more data points in the HI2 FOV explaining the better statistics with 23.33\,\% of data point lying out side of one $\sigma$. For HI2 most of the outliers are clustered at the end of the event, due to the fact the CME front becomes extremely faint and hard to identify (see right of Figure \ref{err3}). 

\section{Results}
	\label{S-results}	
	The velocities of the CME were calculated using two methods, a plane of the sky assumption from A and B, and from the 3-D reconstruction. In both cases this was accomplished by using the first and last data points in the COR2 FOV using $v=h_{max}-h_{min}/t_{max}-t_{min}$ where $h$ and $t$ are the heights and times of the first and last observations. The velocities on larger time/distance scales were also estimated using the first COR2 data point and the last data point (from HI1 or HI2) in the same procedure.
	
\begin{table}
   \caption{The velocities derived for the events using a 2-D (plane-of-sky assumption) and 3-D data. The COR2 velocities are calculated using the first and last data point in the COR2 FOV while the for the average event the first and last data points from the entire event were used. 
   }
   \label{T-complex}
\begin{tabular}{cccccc  l@{.}l l l}     
  \hline                   
Event & Date &  \multicolumn{3}{c}{COR2 Velocities (km\,s$^{-1}$)} & Event $<V>$ (km\,s$^{-1}$)\\
          &         &  A & B & 3-D & 3-D \\
  \hline
1 & 8--13 Oct 07 & $191\pm6$  & $157\pm6$       & $217\pm6$    & $420\pm9$ \\
2 & 25--28 Mar 08 & $863\pm39$ & $1010\pm39$ & $1020\pm39$ & $478\pm20$ \\
3 & 9--10 Apr 088 & $505\pm58$ & $513\pm58$    & $542\pm58$   & $300\pm8$ \\
4 & 9--13 Apr 08 & $110\pm23$ & $218\pm23$    & $189\pm23$    &$317\pm12$ \\
  \hline
\end{tabular}
\end{table}
	
	\subsection{Event 1: 8-13 October 2007}
	\label{SS-event1}
\begin{figure}    
	 	\centerline{\includegraphics[width=0.95\textwidth,clip=]{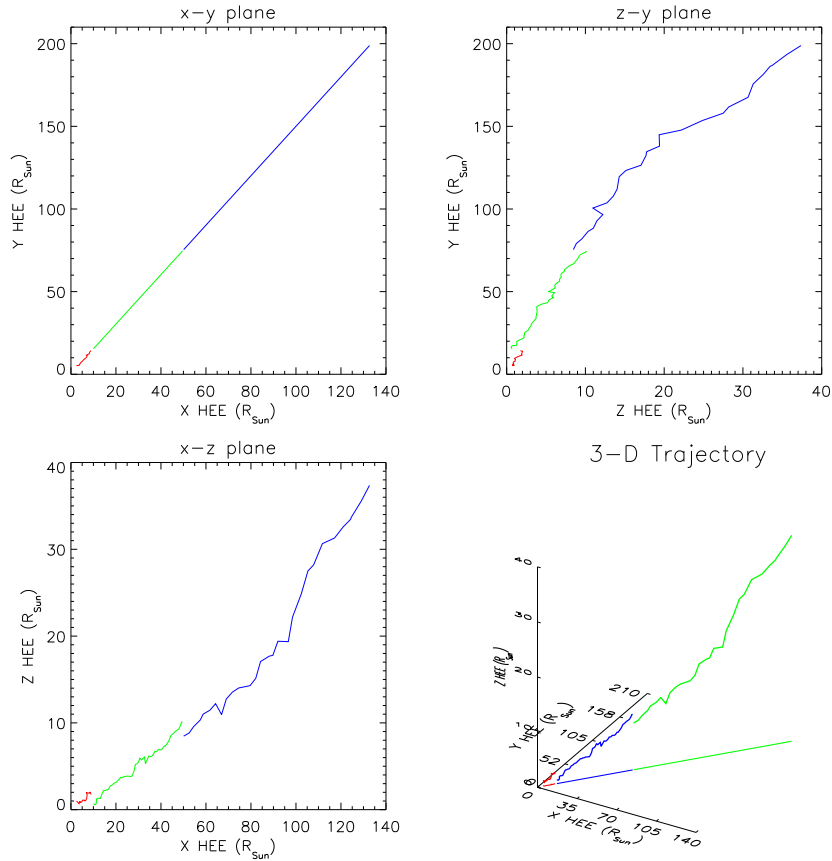}}
		 \caption{CME trajectory for the 8-13 October 2007 event. {\it Top left:} Cut in the $x-y$ plane. {\it Top right:} Cut in the y-z plane. {\it Bottom left:} Cut in the x-z plane. {\it Bottom right:} 3-D trajectory. Note, the axes scales are not the same.}
	   	\label{F-20071009traj}
	\end{figure}

	This event was first observed in COR1 B on 2007 October 8 at 10:15 UT off the west limb, but was extremely faint in images from both spacecraft such that the COR1 data was not used for the 3-D reconstruction (see Figure 3). The first data point used for the 3-D reconstruction was observed at 15:52 UT on the same day by COR2 A and B instruments, the last data point was observed at 04:10 UT on 2007 October 13 by the HI 2 B instrument. From the 3-D data, the launch angle in the $x-y$ plane was found to be $56^{\circ}$ (where $0^{\circ}$ corresponds to an Earth directed CME and $90^{\circ}$ to a CME off the western limb), thus this was a front side event (see Figure \ref{F-20071009traj}). The CME also had an out of ecliptic component of $11^{\circ}$ (where $0^{\circ}$ corresponds to in ecliptic while $90^{\circ}$ corresponds to solar north). The velocity in the COR2 FOV as viewed from A was found to be 191\,km\,s$^{-1}$ from B, 157\,km\,s$^{-1}$, and from the 3-D reconstruction, 217\,km\,s$^{-1}$. The average velocity over the entire event was was found to be 420\,km\,s$^{-1}$. The true height was tracked from 5.7--239.4 R$_{\odot}$ with an estimated error of 0.15 R$_{\odot}$, 0.25 R$_{\odot}$ and 2 R$_{\odot}$ in COR2, HI1 and HI2 respectively. The morphology evolves from a nearly circular loop-like structure in COR1 B and COR2 A and B to a elliptical structure with much larger extent in the vertical than the horizontal direction in HI1 (see Figure \ref{F-20071008mimg}). In HI2, the front evolves from nearly linear to a concave front where the wings/flank of the CME are ahead of the centre, what was originally the CME apex (see Figure \ref{F-20071008mimg}). There is some indication of a multiple loop-like features in COR2 A, HI1 B and HI 2 B (see Figure \ref{F-20071008mimg}) apparent in movies of the event.
	
	%
		
	\subsection{Event 2: 25-28 March 2008}
	\label{SS-event2}
\begin{figure}    
	 	\centerline{\includegraphics[width=0.95\textwidth,clip=]{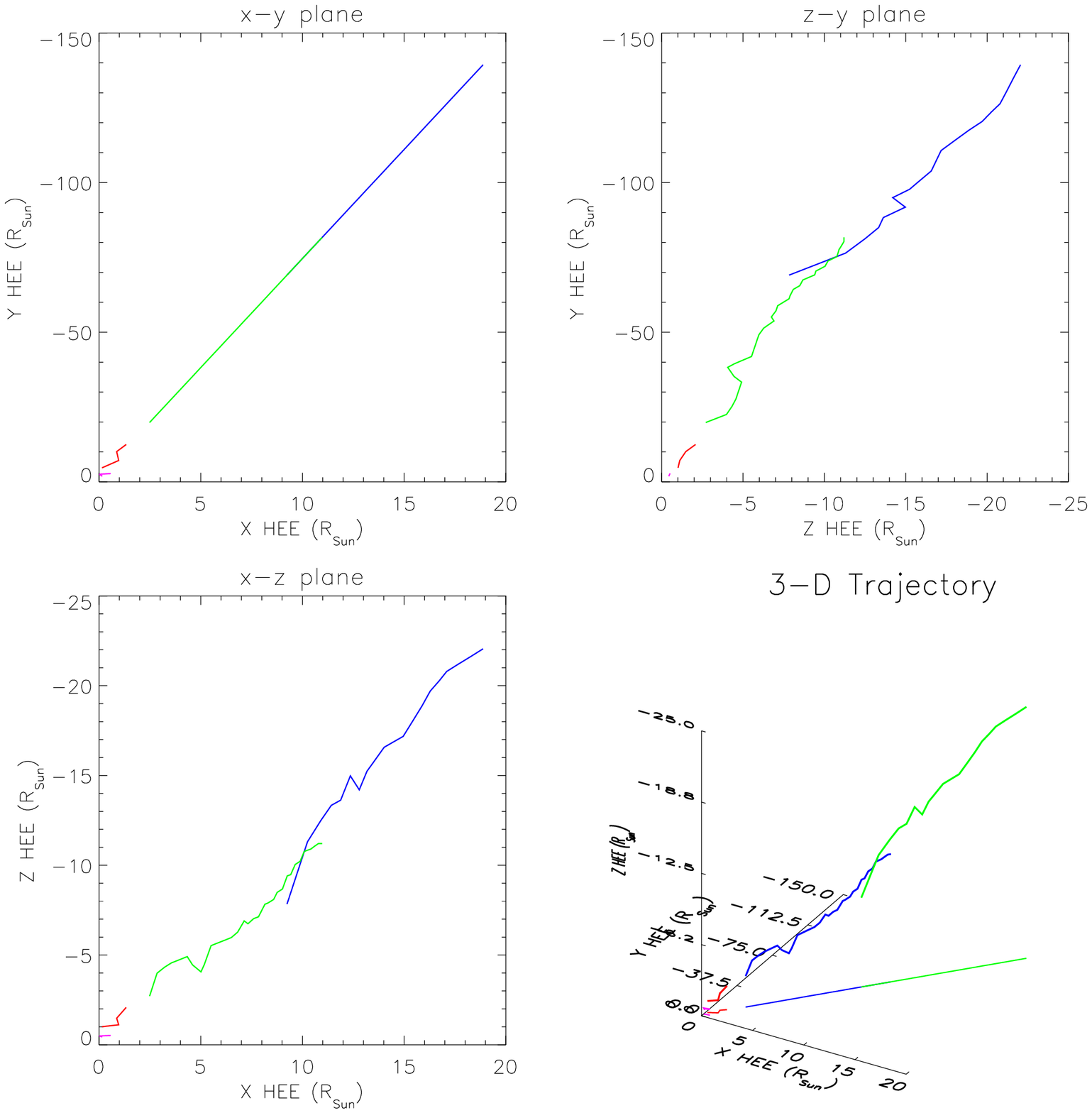}}
		 \caption{CME trajectory for the 25-28 March 2008 event. {\it Top left:} Cut in the $x-y$ plane. {\it Top right:} Cut in the y-z plane. {\it Bottom left:} Cut in the x-z plane. {\it Bottom right:} 3-D trajectory. Note, the axes scales are not the same}
	   	\label{F-20080325traj}
	\end{figure}

The second event was first observed on 2008 March 25 at 18:55 UT in both COR1 A and B off the east limb, the last data point used was observed at 02:09 UT on 2008 March 27. The CME was well observed in COR1 A and B but was somewhat faint and diffuse in COR2. The CME launch angle in the $x-y$ plane was found to be $-74^{\circ}$, which corresponds to a front side event, while the angle with respect to the ecliptic plane was $-22^{\circ}$ (see Figure \ref{F-20080325traj}). The velocity in the COR2 FOV as viewed from A was found to be 863\,km\,s$^{-1}$, from B, 1010\,km\,s$^{-1}$ and from the the 3-D reconstruction, 1020\,km\,s$^{-1}$. The velocity over the entire event was was found to be 478\,km\,s$^{-1}$. The height was tracked from 1.9--139.7\,R$_{\odot}$ with an estimated error of 0.3\,R$_{\odot}$ in HI1 and 0.6\,R$_{\odot}$ in HI2. There were not enough data points in the COR1/2 FOV to estimate the error. The morphologies from both perspectives are very similar. A nearly circular profile consisting of one faint loop-like structure which propagates all the way out into the COR 2 FOV (see Figure \ref{F-20080325mimg}). In the HI1 FOV, there there appears to be a second loop-like feature behind the main front which is expanding at a similar rate as the main front. The CME is tracked until about half way across the HI2 FOV before it became too difficult to identify. The second loop-like structure could be associated with a second flux rope or perhaps prominence material.
	
	%

	\subsection{Event 3: 9-10 April 2008}
	\label{SS-event3}
\begin{figure}    
	 	\centerline{\includegraphics[width=0.95\textwidth,clip=]{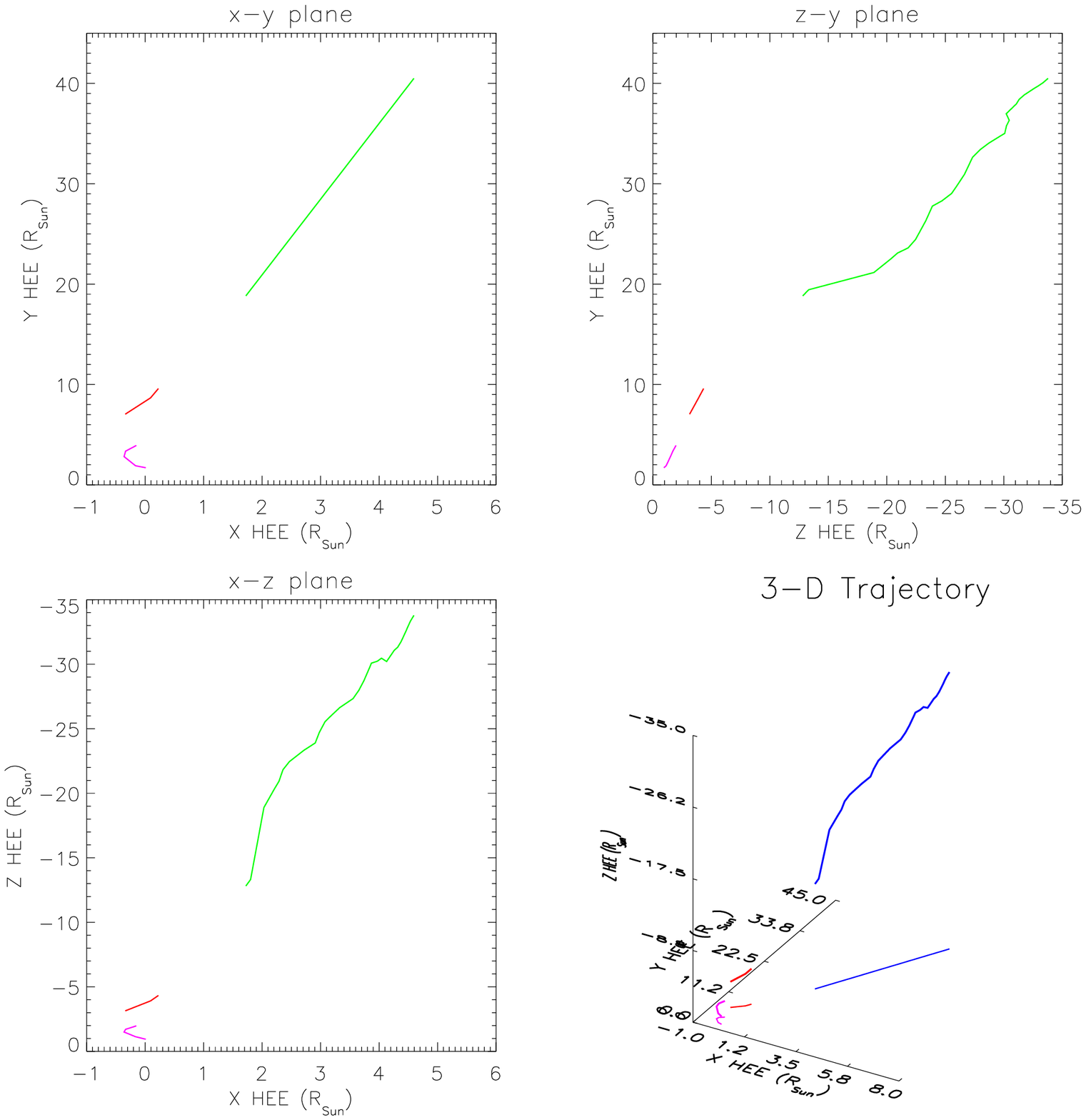}}
		 \caption{CME trajectory for the 9-10 April 2008 event. {\it Top left:} Cut in the $x-y$ plane. {\it Top right:} Cut in the y-z plane. {\it Bottom left:} Cut in the x-z plane. {\it Bottom right:} 3-D trajectory. Note, the axes scales are not the same}
	   	\label{F-20080409traj}
	\end{figure}

This event was first observed on 2008 9 April in both COR1 A and B at 10:05 UT off the south west limb (see Figure 5). The last data point was observed at 15:29 UT on the 2008 April 10 in HI1 B. The CME was well observed in A and B images from both COR1 and COR2. The CME launch angle in the x-y plane was found to be $-71^{\circ}$Ò, which corresponds to a front side event, while out of the ecliptic plane $-60^{\circ}$. Due to the large out of ecliptic angle the CME does not pass through the FOV of HI2 B. The velocity in the COR2 FOV as viewed from A was found to be 505\,km\,s$^{-1}$, from B, 513\,km\,s$^{-1}$ and from the 3-D reconstruction, 542\,km\,s$^{-1}$. The average velocity over the entire event was was found to be 300\,km\,s$^{-1}$.  The height was tracked from 1.9--50.4\,R$_{\odot}$ with an estimated error of 0.5\,R$_{\odot}$ in HI1. There were not enough data points in the COR1/2 to estimate the errors. The CME appears to be narrower in COR1 A than in COR1 B. When the CME enters the COR2 FOV it appears the same size in A and B (see Figure \ref{F-20080409mimg}). The CME has a more elliptical shape in HI1 and continues to propagate at a steep angle. There is no obvious trend of the trajectory back towards the ecliptic plane which one may have expected (see Figure \ref{F-20080409traj}).
	%
		
	\subsection{Event 4: 9-13 April 2008}
	\label{SS-event4}
\begin{figure}    
	 	\centerline{\includegraphics[width=0.95\textwidth,clip=]{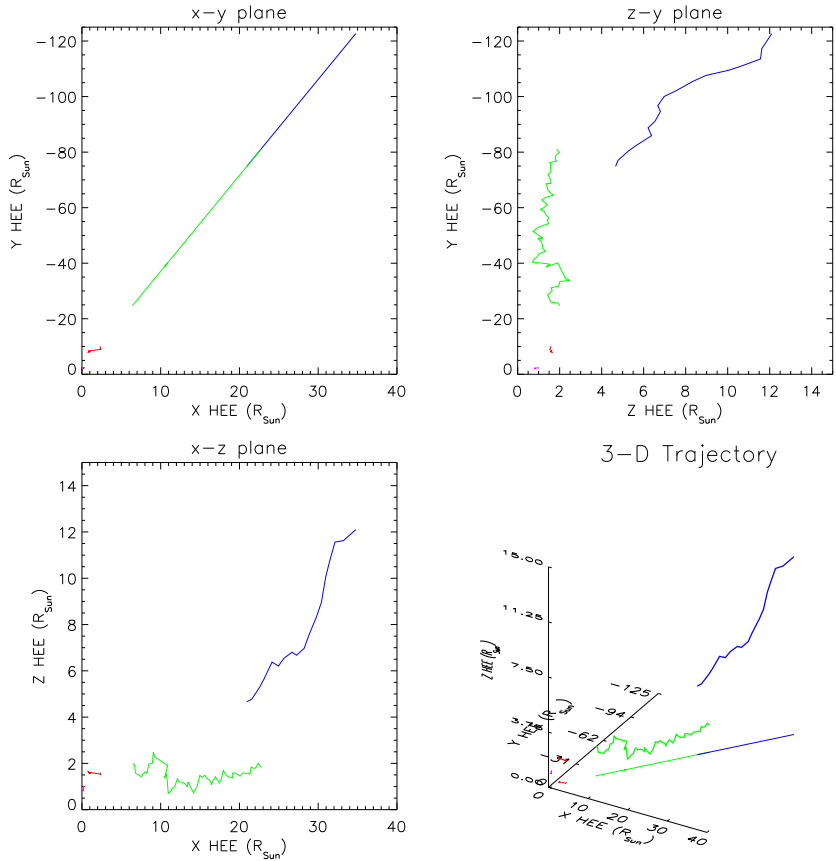}}
		 \caption{CME trajectory for the 9-13 April 2008 event. {\it Top left:} Cut in the $x-y$ plane. {\it Top right:} Cut in the y-z plane. {\it Bottom left:} Cut in the x-z plane. {\it Bottom right:} 3-D trajectory. Note, the axes scales are not the same}
	   	\label{F-20080410traj}
	\end{figure}

This event was first observed off the the east limb on the 9 April 2008 at 15:05 UT by both COR1 A and B instrument (see Figure 6). The last data point was observed by HI2 A  at 00:09 UT on 2008 April 13. The CME was extremely faint in both A and B COR1 but a small number of data points were able to be recorded. The CME was reasonably clear in COR2 A but very faint in B. As a result, the CME front was not visible and for this event a dark linear feature was tracked instead (see Figure \ref{F-20080410mimg}). The CME launch angle in the x-y plane was found to be $-51^{\circ}$ which corresponds to a front side event, while out of the ecliptic plane $16^{\circ}$ (see Figure \ref{F-20080410traj}). The velocity in the COR2 FOV as viewed from A was found to be 110\,km\,s$^{-1}$, from B, 218\,km\,s$^{-1}$ and from the 3-D reconstruction, 189\,km\,s$^{-1}$. The average velocity over the entire event was was found to be 317\,km\,s$^{-1}$. The CMEÕs height was tracked from 2.2--128.0\,R$_{\odot}$ with an estimated error of 0.8 R\,$_{\odot}$ in HI1 and 1.1\,R$_{\odot}$ in HI2 (there were not enough data points in the COR1/2 to estimate the errors). The CME appears very narrow in COR2 B images which could be a loop-like structure with it's axis nearly perpendicular to the plane of sky. The CME has an elliptical shape in HI1 with some material in the centre (see Figure \ref{F-20080410mimg}). This shape is held into HI2 where the CME is lost as it crosses in front of the Milky Way (where the background subtraction did not perform well).

%
	
\section{Discussion and Conclusions}
	\label{S-disc}
We have reconstructed the 3-D trajectories of CMEs in the inner Heliosphere using STEREO/SECCHI observations. These 3-D reconstructions give the true positions of the CMEs studied in the range between $\sim$2--240\,R$_{\odot}$ as a function of time. The COR1/2 data indicate that a radial propagation model is appropriate once the CME is above a few solar radii. This is in agreement with what is expected as the initial acceleration acting on a CME falls off steeply with increasing radius (especially the component which would act perpendicular to the propagation direction). We have also shown that the velocities derived from a single perspective can be at odds with values derived from a 3-D reconstruction. Finally, we have shown that CMEs undergo acceleration in the Heliosphere as the velocities in the COR2 FOV can be very different to those in the HI FOV.

There are a number of outstanding issues with the reconstructions. Some of these are the result of the reconstruction method, while others are related to the data used or its interpretation. The stereoscopy method requires that we can identify the same feature in both images and this is extremely difficult in some cases where the CME is very faint or has a very different appearance from each spacecraft. For example, the COR1 data for first event was so faint the CME could not be identified in the A images (see Figure \ref{F-20071008mimg}) and in both the third and fourth events, the identification of the same feature was extremely difficult as the CME has a different appearance in each image (see Figures \ref{F-20080409mimg} and \ref{F-20080410mimg}). This is evident in the reconstructions as they are less contiguous and there are large changes in CME trajectories. These large changes are unphysical as there is no source for such large forces acting in a direction transverse to the propagation in the Heliosphere.

The different sensitivities and light rejection levels of the each instrument means that CMEs may appear larger or smaller due to instrumental effects. This leads to the discontinuities of the first event. Furthermore, as the front of the CME is a line-of-sight integration through the CME (flux rope or magnetic bubble), it will appear different from each spacecraft. This is especially important in the HI FOV as CMEs can expand up to large sizes. This means that the positions, and thus heights derived, will always be an upper limit on the actual values. If a model shape for CMEs is used, such as an ellipse or circle, we can estimate the size of this effect. We await an event which is directed at one of the {\it in situ} instruments and can be reconstructed as this would allow for comparison of our reconstruction with the actual CME's path. Finally, the coordinate transforms and geometry used in the reconstruction are not trivial and need to be taken into account (see the appendix for details). 

A number of interesting observations arise from these data. The difference in the velocities of the CMEs in the COR and HI FOV, the possible multi-loop nature of two of the CMEs, the complex structure seen at the rear of the CMEs (especially in HI1) and finally the distortion of the CME front. In Table 1 we can see that for the fast events (Events 2 and 3) the velocities in the COR FOV are higher than the event average, and for slow events (Events 1 and 4) the opposite is true. This tendency for fast CMEs to be decelerated and slow CME to be accelerated could be caused by an interaction between the CME and the solar wind. In all cases, the velocity changes to approach a more typical solar wind velocity. The mulit-loop nature is visible in Events 1 and 2 (see Figures \ref{F-20071008mimg} and \ref{F-20080325mimg}). These could be attributed to a second flux rope or possibly due to prominence material. These features are only visible because of the unique features of the HI1 and HI2 instruments. Most of the events show complex structure at the rear of the CME. This is very apparent in Figures \ref{F-20071008mimg}, \ref{F-20080409mimg} and \ref{F-20080410mimg}. This type of structure could be interpreted as the rear part of a flux rope if the flux rope axis was into or out of the image plane. Finally, the distortion of the CME front in Event 1 as it progresses through the Heliosphere could be caused by a number of processes. This may be a Thomson scattering effect i.e., we are seeing different parts of the CME from different regions along the lines of sight and it only appears as though the CME front is distorted with the flanks ahead of the nose. The other possibility is that as the CME has expanded to an extremely large size at this point, the nose and flanks are experiencing very different ambient solar wind conditions. As such, the CME flank may be in the high speed solar wind region (possibly accelerated by aerodynamic drag), or the nose may be in a relatively density enhanced region due to the heliospheric current sheet. This implies, it is experiencing more drag than the CME flanks.

These reconstructions are the first steps towards the detailed study of CME kinematics in the Heliosphere. Using these trajectories, the full 3-D CME kinematics which will not be subject to projection effects can be derived. Combining these with CME mass estimates will provide all the information needed to study the forces acting on CMEs from $\sim$2\,R$_{\odot}$ to beyond 1\,AU. This will allow a comparison of the derived forces with those predicted by the various models (flux rope, snow plough and aerodynamic drag). 



 \begin{acks}
This work is supported by a grant from Science Foundation Ireland's Research Frontiers Programme. JMA is currently funded by a Marie Curie Intra-European Fellowship. We would like to thank the STEREO/SECCHI consortium for providing open access to their data and technical support. The authors would also like to thank the HI team and in particular Danielle Bewsher. 
\end{acks}

\appendix

\section{Coordinate transforms in COR1 and COR2}
	\label{SS-reconcor12}
	
The derivation of Helioprojective-Cartesian coordinates from pixel coordinates is a multi step procedure. In the first step, pixel coordinates are transformed to intermediate coordinates i.e., they are converted into the relevant units (meters, degrees, arcsec) but they are not adjusted for the reference point of the observations or projection/geometrical effects. These intermediate coordinates are then transformed by the various possible projections (TAN, AZP, etc) and rotated into celestial coordinates. In the case of COR1/COR2, the small Sun centred FOV means that the full spherical TAN or Gnomonic projection could be approximated using the the small angle formula. The conversion from pixel to intermediate coordinates is given by

\begin{equation}
	x_{i} = s_{i}\sum_{j=1}^{N}m_{ij}\left (  p_{j} - r_{j}\right )
	\label{Eq-intercoords}
 \end{equation}where $p_{i}$ refers to pixel coordinates, $r_{j}$ is the reference pixel, $m_{ij}$ is a linear transformation matrix and $s_{i}$ is a scale function, with subscript $i$ referring to pixel axes and subscript $j$ referring to coordinate axis \cite{Calabretta:2002p1915}. 
As a direct result of the small angle approximation we can readily convert to full Helioprojective-Cartesian by adding the reference co-ordinate $c_{i}$ to the intermediate co-ordinates as $x_{i} = x_{i} + c_{i}$ (where $x_{0}$ and $x_{1}$ correspond to $\theta_{x}$ and $\theta_{y}$). A more intuitive way of describing this coordinate system is given below,
   \begin{eqnarray}   
		\theta_{x} \approx \left ( \frac{180^{\circ}}{\pi} \right ) \frac{x}{D_{\odot}} \approx  \left ( \frac{180^{\circ}}{\pi} \right ) \frac{x}{d}
		  \label{Eq-smallangx}
		 \\      
		 \theta_{y} \approx \left ( \frac{180^{\circ}}{\pi} \right ) \frac{y}{D_{\odot}} \approx  \left ( \frac{180^{\circ}}{\pi} \right ) \frac{y}{d}
                       \label{Eq-smallangy}    
   \end{eqnarray} where $D_{\odot}$ is the Sun-observer distance and $d$ is the observer feature distance. A Solarsoft routine ${\it scc\_measure.pro}$ was used to find the 3-D coordinates as it implements the method above. This routine produces a list of 3-D coordinates in Stonyhurst Heliographic Coordinates which can be transformed into Heliocentric Earth Ecliptic (HEE), which is the coordinate system used for all our reconstructions \cite{Thompson:2006p44}.

	\section{Coordinate transforms in HI1 and HI2}
 		\label{SS-reconhi12}
		In the HI FOV the derivation of Helioprojective-Cartesian coordinates from pixel coordinates is more complex. The optical system of HI is described by the {\it Slant} Zenithal (Azimuthal) Projection (AZP) model. The first step is to convert to intermediate coordinates as described above in Section \ref{SS-reconcor12} and by Equation 5. The transformation between the intermediate coordinates and Helioprojective-Cartesian is much more complex as a full {\it Slant} (AZP) is required in which
 
   	\begin{equation}   
		\phi = \tan^{-1} \left ( x / -y \cos \left ( \gamma \right ) \right )
		  \label{Eq-azpphi}
	\end{equation}
	\begin{equation}
		\theta  = 
		 \begin{cases} 
		 	\psi - \omega \\
		 	\psi + \omega + 180^{\circ}
		 \end{cases}
		\label{Eq-azptheta}
	\end{equation} where
	\begin{equation}
		 \psi = \tan^{-1} \left ( 1/\rho \right )
		 \label{Eq-azppsi}
	\end{equation}
	\begin{equation}
		 \omega = sin^{-1} \left ( \frac{\rho \mu}{\sqrt{\rho^{2}+1} } \right )
		 \label{Eq-azpomega}
	\end{equation}
	\begin{equation}
		\rho = \frac{ R }{ \frac{ 180^{\circ }}{ \pi } \left ( \mu +1 \right ) + y \sin \left ( \gamma \right )}
		\label{Eq-azprho}
	\end{equation}
	\begin{equation}
		 R =  \sqrt{x^{2} + \left ( y \cos \left ( \gamma \right ) \right )^{2} }
		 \label{Eq-azprtheta}    
	\end{equation}	 where $\gamma$ is the look-angle or the angle between a line to the centre of the coordinate system and the centre of the camera optical axis and $\mu$ is the distortion parameter. Once $\phi$ and $\theta$ are calculated they can then be rotated into celestial coordinates to arrive at Helioprojective-Cartesian coordinates ($\theta_{x}$ and $\theta_{y}$).

%
%
 \bibliographystyle{spr-mp-sola-cnd}
 \bibliography{newbib}  

\end{article} 
\end{document}